\documentclass[12pt]{article}
\usepackage{amssymb,epsfig}

\def\dmatm{\Delta m_{@}^2}
\def\dmsun{\Delta m_{\odot}^2}
\def\tg2thsun{\tan^2\theta_{\odot}}
\def\eV{\mbox{\rm eV}}
\def\calN{{\cal N}}
\def\calO{{\cal O}}
\def\TT{^{\rm T}}
\def\diag{{\rm diag}}

\def\UMNS{U_{\rm MNS}}
\begin{document}
\begin{titlepage}
\title{
\vspace{2cm}
A hidden hierarchy of neutrino masses}
\author{M.~Je\.zabek$^{a,b}$
and
P.~Urban$^{a,c}$
\\ \\ \\ \small
$a$ Institute of Nuclear Physics,
Kawiory 26a, PL-30055 Cracow, Poland
\\   \small 
 $b$ Institute of Physics, University of 
      Silesia, \\[-5pt]  \small
     Uniwersytecka 4, PL-40007 Katowice, Poland\\ 
\small
$c$ Institut f\"ur Theoretische Teilchenphysik,
Universit\"at Karlsruhe,\\  [-5pt]
\small
D-76128 Karlsruhe, Germany
}
\date{}
\maketitle
\thispagestyle{empty}
\vspace{-4.5truein}
\begin{flushright}
{hep-ph/0206080}\\
{June 2002}
\end{flushright}
\vspace{4.0truein}
\begin{abstract}
The Dirac masses of neutrinos can exhibit a strong hierarchy even if
the Majorana masses of the right-handed neutrinos are degenerate and
the hierarchy of the mass scales governing the oscillations of solar
and atmospheric neutrinos is rather small. This phenomenon results
from the see-saw mechanism and the algebraic structure of the effective
mass operator for the active neutrinos. The large hierarchy of the
Dirac masses is drastically modified by a symmetric unitary matrix $R$
acting in the flavor space.
A realistic pattern of neutrino masses and mixing
is obtained. Maximal mixing for atmospheric neutrinos is attributed to
the charged lepton sector. Large mixing in solar neutrino oscillations
is due to the neutrino sector. Small $U_{e3}$ is a natural consequence
of the model. The masses of the active neutrinos are given by
$\mu_3\approx\sqrt{\Delta m_{@}^2}$ and $\mu_1/\mu_2\approx
\tan^2\theta_\odot$. 
\end{abstract}
\end{titlepage}

\section{Introduction}  
After four decades of heroic experimental \cite{Homestake,SK-solar,SAGE,
GALLEX,SNO} and theoretical \cite{BPB} efforts, 
see \cite{bgg,pdg00} and references therein, 
the problems of solar neutrinos have finally been
solved. A unique solution of these problems exists corresponding to
LMA MSW flavor oscillations of active neutrinos. By active neutrinos we
understand those experimentally observed, $\nu_e$, $\nu_\mu$ and 
$\nu_\tau$. After recent results from
SuperKamiokande \cite{SK-solar} and SNO \cite{SNO} all other solutions 
of the solar neutrino problems are not allowed at $3\sigma$ 
level \cite{phenosol}. 
Simultaneously it becomes more and more
clear that the oscillations of atmospheric neutrinos are due to
$\nu_{\mu}\rightleftarrows \nu_{\tau}$ transitions 
\cite{SK-atmo,K2K}. 
The third important piece of information is the CHOOZ limit \cite{CHOOZ} 
indicating that
the element $U_{e3}$ of the Maki-Nakagawa-Sakata (MNS) lepton mixing
matrix \cite{MNS} is small. In this article we show that all these
experimental facts can be nicely accommodated by assuming the see-saw
mechanism \cite{seesaw} and a large hierarchy of the Dirac masses for
three neutrinos. One can consider the results
presented in the present paper as a realistic realization of the ideas
proposed in a recent paper by one of us \cite{j02}. It should be
mentioned that there is no room in this model to explain LSND results
\cite{LSND} in terms of oscillations. This model cannot also
accommodate the large value of the Majorana mass parameter
$\left<m_{\nu_{e}}\right>$ which follows from a recent interpretation
\cite{klapdor} of the Heidelberg-Moscow data on neutrinoless
double beta decays of $^{76}$Ge \cite{heidelberg-moscow}. 

Let us start with a brief summary of the ideas and results presented in
\cite{j02}. It has been shown in \cite{j02} that a rather small
hierarchy of the observed low energy masses of the active neutrinos can
follow from a large hierarchy at the more fundamental level of the Dirac
masses. This can happen even in the case when the Majorana masses of the
right-handed neutrinos are degenerate. Therefore large hierarchies in the
Dirac masses of all other fundamental fermions, i.e. up quarks, down
quarks and charged leptons, can be accompanied by a large hierarchy of
the Dirac masses for the neutrinos. This large hierarchy is drastically
modified by a symmetric unitary operator $R$ related to unitary
transformations of the right-handed neutrinos. In some sense the
underlying hierarchy of the Dirac masses is hidden and only mildly
reflected in the effective low energy masses. This phenomenon is due to
the see-saw mechanism and the algebraic structure of the low energy
effective mass operator $\cal N$ describing the masses of the active 
neutrinos.
 
Therefore, in the lepton sector there are two operators acting in the
flavor space which affect the low energy physics in a most important
way. One is the MNS lepton mixing matrix $\UMNS$ \cite{MNS} which
plays a role similar to that of the Cabibbo-Kobayashi-Maskawa mixing
matrix for quarks \cite{CKM}. The other is the matrix $R$ which is
imprinted in the observable mass spectrum of the active neutrinos. Up to
our knowledge this remarkable matrix $R$ and its role in the low energy
physics of neutrinos was not discussed in the literature before
\cite{j02}. In \cite{j02} a form of $R$ has been considered leading to a
significant reduction of the mass hierarchy and a form of $\UMNS$ nicely
accommodating the small value of its $U_{e3}$ element. It has also been
argued that the mass ratio for the two lighter active neutrinos is given
by $\tan^2\theta_\odot$ with $\theta_\odot$ being the mixing angle for
solar neutrinos. In the present paper we show that for strongly 
hierarchical Dirac masses the form of $R$ considered in \cite{j02} 
is up to complex phase factors the only one leading to acceptable 
low energy mass spectrum and lepton mixing matrix. We show also that 
the smallness of $U_{e3}$ and the mass relations are preserved in the 
realistic model.


\section{Hidden hierarchy}

In this section the see-saw mechanism for three families of leptons is
discussed. We repeat arguments given in \cite{j02} and show that a small
hierarchy of low energy masses can be obtained from a large hierarchy
in the Dirac masses. We show that the observed small hierarchy for the
active neutrinos fixes the form of the matrix $R$ mentioned in the
Introduction up to some complex phase factors and small sub-leading terms. 
%
%
The mass matrix for the charged leptons can be written as
\begin{equation}\label{eq:L}
L = V_R \; {\rm diag}(m_e,m_\mu,m_\tau)\; V_L \equiv V_R \;m^{(l)}\; V_L
\label{eq:1}
\end{equation}
The matrix $V_R$ multiplying $m^{(l)}$ from the left side can be made
equal to one by an appropriate redefinition of the right-handed charged
leptons. This has no observable consequences because at our low energies
only left-handed weak charged currents can be studied. The corresponding
Dirac mass matrix for the neutrinos is
\begin{equation}\label{eq:N}
N = U_R\; m^{(\nu)}\; U_L
\end{equation}
with
\begin{equation}\label{eq:mnu}
m^{(\nu)}={\rm diag}(m_1,m_2,m_3) \quad .
\end{equation}
We choose the reference frame in the flavor space such that the Majorana
mass matrix of the right-handed neutrinos is diagonal. In general in this
frame $M_R$ is described by its three different eigenvalues. However, as
we demonstrate in the following, even taking the simplest case of 
\begin{equation}\label{eq:MRequal1}
M_R=M\cdot {\mathbf 1}
\end{equation}
we are able to build a phenomenologically successful model.
The Majorana masses of the right-handed neutrinos are equal to $M$,
which is huge. This leads to the masses of active neutrinos much smaller
than the masses of other fundamental fermions (leptons and quarks). 

Of course there is an additional freedom for the right-handed
Majorana masses not equal to each other but one can describe the
neutrino oscillation data without any problem assuming that they are
all equal. In our present purely phenomenological approach we do not
have at our disposal a sufficient amount of information to study
the mass spectrum of $M_R$. This may be possible in a more complete
theory of flavor, but in this work we assume degenerate Majorana masses.
It should be noted, however, that the arguments which we present can
be trivially generalized and applied to the case of non-equal right-handed
Majorana masses.

The form of the matrix $M_R$ depends on the reference frame in a non-trivial
way. In the frame employed in the present paper this matrix is assumed to
be proportional to the unit matrix. However it may have a different form in
another reference frame. In particular it need not be proportional to the 
unit matrix in the frame where the matrix of neutrino Dirac masses is
diagonal and so exhibits a strong hierarchy. 

Once the reference frame in the flavor space is fixed by the condition
that $M_R$ is diagonal the 
matrix $U_R$ in eq.(\ref{eq:N}) cannot be gauged away by a redefinition
of the right-handed neutrino fields. Let us assume that $m_1,m_2$, and 
$m_3$ are hierarchical 
\begin{equation}
m_1 \ll m_2 \ll m_3
\label{eq:hierarchical}
\end{equation}
which means that the Dirac masses of the neutrinos exhibit
the same strongly ordered hierarchical pattern as the Dirac masses
of all other fundamental fermions. The mass spectrum of the
active neutrinos is dictated by an effective mass operator $\calN$ of
dimension five
\begin{equation}\label{eq:defN}
\calN = N^{\rm T}M_R^{-1}N=U_L^{\rm T}
m^{(\nu){\rm T}}U_R^{\rm T}M_R^{-1}U_R m^{(\nu)}U_L.
\end{equation}
The algebraic structure of $\calN$ implies that the resulting mass 
spectrum is extremely sensitive to the form of 
\begin{equation}
U_R^{\rm T} M_R^{-1} U_R  = {1\over M} U_R^{\rm T} U_R
\label{eq:7}
\end{equation}
where the simplifying assumption (\ref{eq:MRequal1}) has been used.
The mass spectrum obtained from the matrix
$\calN$ in eq.(\ref{eq:defN}) is now 
seen to depend crucially on the following matrix $R$,
\begin{equation}\label{eq:defR}
R=U_R^{\rm T}U_R,
\end{equation}
which is symmetric and unitary. It is remarkable that to some
extent the condition  (\ref{eq:MRequal1}) also fixes the reference frame 
in the flavor space. Unitary transformations of the right-handed fields 
are in general complex and cannot be gauged away even if $M_R$ is 
proportional to the unit matrix. 
The matrix $R$ can drastically reduce the hierarchy of the mass spectrum
for the active neutrinos. So, $R$ is observable, in principle at least, if
a large hierarchy of the Dirac masses is the common feature of all quarks
and leptons. In this sense $R$ is a physical object which is imprinted in 
low energy physical quantities, namely the masses of the active neutrinos.
Unlike the quark sector with its Cabibbo-Kobayashi-Maskawa mixing 
matrix \cite{CKM} the lepton sector has therefore two important
matrices in the flavor  
space. One is the lepton mixing matrix $U_{\rm MNS}$ \cite{MNS} 
which affects the form of the weak charged current. 
Another is the matrix $R$ defined in eq.(\ref{eq:defR}). 
$R$ affects the form of $U_{\rm MNS}$. Moreover, it is also reflected
in the low energy neutrino mass spectrum. In our phenomenological approach
we use the experimental input to fix the form of $R$. One may hope that
this is a first step towards an underlying theory of flavor. 

What can be said about the matrix $R$? In \cite{j02} it has been shown
that its element $(R)_{33}$ must be equal to zero at the leading order.
By this we mean that $(R)_{33}$ cannot be a number of order 1. It can be
a small number suppressed by some power of the mass ratio of, say, $m_1$
and $m_3$. Perhaps the easiest way to demonstrate that it must be so is
to take $R=1$ which is representative for the whole class of matrices with
$(R)_{33} = {\cal O}(1)$. Then for the hierarchical mass spectrum
(\ref{eq:hierarchical}) a much stronger hierarchy is obtained for the
masses of the active neutrinos
\begin{equation}\label{ActiveHierarchy}
\mu_1=\frac{m_1^2}{M} \ll \mu_2 = \frac{m_2^2}{M} \ll \mu_3 =
\frac{m_3^2}{M}\quad .
\end{equation}
It should be stressed that in eq.(\ref{ActiveHierarchy}) much greater
means a few orders of magnitude rather than factors below 10.
Then for the ratio
\begin{equation}
\rho = \frac{\dmsun}{\dmatm} = \frac{\mu_2^2-\mu_1^2}{\mu_3^2-\mu_2^2}
\label{eq:rho}
\end{equation}
values between $10^{-4}$ and $10^{-8}$ are obtained if the mass ratio
$m_3/m_2$ is taken of the order of the corresponding mass ratios for
other fundamental fermions, i.e. $m_b/m_s \sim 30$, 
$m_\tau/m_\mu \approx 17$ or $m_t/m_c \sim 100$. These values of $\rho$
are two to six orders smaller than the experimental value
\begin{equation}
\rho_{exp} \approx \frac{5\cdot 10^{-5} \eV^2}{2.5 \cdot 10^{-3}
  \eV^2}=2\cdot 10^{-2}.
\label{rhoexp}
\end{equation}
In the above equation the value of $\dmatm$ from a combined fit to
SuperKamiokande and K2K data \cite{phenoatm} 
and a recent SNO result \cite{SNO}
for $\dmsun$ are used. Before SNO, much smaller values of $\dmsun$ 
were also allowed for the oscillations of the solar neutrinos.
As we see, if $(R)_{33}$ is not equal to zero, then an
unacceptably large hierarchy appears in the mass spectrum of the
active neutrinos. This means that $(R)_{33}=0$ must be assumed.
 
What about $(R)_{23} = (R)_{32}$? It turns out that the resulting
mass spectrum for the active neutrinos is acceptable from the phenomenological
point of view if $(R)_{23} = {\cal O}(1)$ is assumed. This spectrum 
corresponds to the case of the so-called inverted hierarchy. However in the
following section it will be seen that the resulting structure of
the lepton mixing matrix does not resemble the experimentally observed one. 
More precisely, one can obtain a realistic mixing matrix assuming a rather
artificial form of the charged lepton contribution $V_L$ and some sort of
conspiracy and accidental cancellations. For esthetic reasons we do
not consider such a situation acceptable and conclude that up to small
sub-leading terms  $(R)_{23}=0$ must be assumed.

The only remaining case is $R_{33}=R_{23}=0$ which implies
\begin{equation}\label{eq:R0}
R= \left( 
\matrix{ 0 & 0 & \exp i\phi_1 \cr
         0 & \exp i\phi_2 & 0 \cr
         \exp i\phi_1 & 0 & 0 \cr }\right)
\end{equation}
The complex phase factors in eq.(\ref{eq:R0}) can be of crucial
importance for lepton number violating processes like neutrinoless
double beta decays. However, these phase factors do not affect our 
discussion which concentrates on neutrino oscillations. So, for the 
sake of simplicity, in the following considerations we take the same
form of $R$ as in \cite{j02}:
\begin{equation}\label{eq:R}
R= \left( 
\matrix{ 0 & 0 & 1 \cr
         0 & 1 & 0 \cr
         1 & 0 & 0 \cr }\right)
\equiv P_{13}
\end{equation}
It turns out that for strongly hierarchical Dirac masses eq.(\ref{eq:R}) 
is a necessary condition for a realistic mixing and mass spectrum. 
Therefore we assume that some symmetry underlying flavor
dynamics forces $U_R$ to fulfill eq.(\ref{eq:R}). 

Unlike $V_R$ in the charged lepton sector which at low energies can be 
gauged away, the matrix $U_R$ is imprinted in low energy physics and its 
structure affects the mass spectrum of the active neutrinos in a most
spectacular way. The general solution of eq.(\ref{eq:R}) is
\begin{equation}\label{URsolved}
U_R={\cal R}(\alpha,\beta,\gamma)\frac{1}{\sqrt{3}}\left(
\matrix{ \omega   &  1 & \omega^* \cr
         1        &  1 & 1        \cr
         \omega^* &  1 & \omega }\right)
\end{equation}
where ${\cal R}(\alpha,\beta,\gamma)$ is an arbitrary 3-dimensional
rotation and $\omega=\exp(2\pi i/3)$. The particular solution in
(\ref{URsolved}) has been chosen to be a symmetric matrix. One may
speculate if this interesting matrix plays some role in the theory of
flavor.


\section{Lepton mixing matrix}

In this section we study the MNS mixing matrix. Following \cite{j02}
we show that for the matrix $R$ given in eq.(\ref{eq:R}) the so-called
bi-maximal mixing is obtained \cite{bi-maxim}. It is known \cite{phenosol}
that after SNO \cite{SNO} the bi-maximal mixing is excluded as a realistic
description of the data on neutrino oscillations. Moreover, the mass spectrum
of the active neutrinos following from eq.(\ref{eq:R}) is also not realistic 
because two masses are degenerate implying $\dmsun=0$. Both problems are 
cured in the following section by adding appropriate sub-leading terms.
Importantly, the key observation which we present here survives in the
realistic model. We also show that
assuming $(R)_{23} = {\cal} O(1)$ one obtains another structure of $U_{MNS}$
which cannot be made realistic without very artificial assumptions. 

The lepton mixing matrix $U_{MNS}$ relates two sets of neutrino
eigenstates: gauge interaction eigenstates $\nu_e,\nu_\mu$ and
$\nu_\tau$ and mass eigenstates $\nu_1,\nu_2,\nu_3$ of masses
$\mu_1,\mu_2,$ and $\mu_3$:
\begin{equation}
\left( \matrix{ \nu_e    \cr
                \nu_\mu  \cr
                \nu_\tau \cr} \right) =
\left( \matrix{ 
 U_{e1} & U_{e2}  & U_{e3} \cr
 U_{\mu 1} & U_{\mu 2}  &  U_{\mu 3}   \cr
 U_{\tau 1}  & U_{\tau 2}  & U_{\tau 3}   \cr } \right)
\left( \matrix{ \nu_1    \cr
                \nu_2  \cr
                \nu_3 \cr} \right) =
U_{\rm MNS} \left( \matrix{ \nu_1    \cr
                \nu_2  \cr
                \nu_3 \cr} \right) .
\label{eq:UMNS}
\end{equation}
Let $\calO$ be a unitary matrix such that 
\begin{equation}
\calO\TT \calN \calO = \diag(\mu_1,\mu_2,\mu_3) .
\label{eq:calO}
\end{equation}
Eq.(\ref{eq:L}) implies that $M_L^2=L^\dagger L$ is diagonalized by
$V_L$, i.e. $V_L M^2_L V_L^\dagger =
\diag(m_e^2,m_\mu^2,m_\tau^2)$.
Then from eqs.(\ref{eq:defN})-(\ref{eq:defR})
one derives
\begin{equation}\label{UMNSVLO}
U_{\rm MNS}=V_L \calO = V_L U_L^{-1} \calO'
\end{equation}
where the unitary matrix $\calO'$ is such that
\begin{equation}
\frac{1}{M} {\calO'} \TT {m^{(\nu)}}\TT\; R\;  m^{(\nu)}
\calO'=\diag(\mu_1,\mu_2,\mu_3)
\label{eq:18}
\end{equation}
with $R = P_{13}$, c.f. eq.(\ref{eq:R}).
For $m^{(\nu)}$ given in (\ref{eq:mnu}) one obtains,
see the following section for a step by step derivation,
\begin{equation}
{O^\prime} = P_{23} U_{12}\left( {\pi/ 4}\right)
\label{eq:calOprime}
\end{equation}
with
\begin{equation}
P_{23} = P_{23}\TT =
    \left( \matrix{ 1  &   0  &   0  \cr
                    0  &   0  &   1  \cr
                    0  &   1  &   0  \cr } \right)
\label{eq:P23}
\end{equation}
and
\begin{equation}
U_{12}\left( {\pi/ 4}\right) =
\left( \matrix{ 
{i\over\sqrt{2}}  &  {1\over\sqrt{2}} &   0  \cr
-{i\over\sqrt{2}}  &  {1\over\sqrt{2}} &   0  \cr
                    0  &   0  &   1  \cr } \right)
\label{eq:U12pi4}
\end{equation}
The lepton mixing matrix is, cf.(\ref{UMNSVLO}),
\begin{equation}\label{UMNSVUPUpi4}
U_{\rm MNS}=V_L U_L^{-1} P_{23} U_{12}\left(\pi/4\right).
\end{equation}
The presence of $P_{23}$ in eq.(\ref{UMNSVUPUpi4}) may be considered very
embarrassing. Multiplying any matrix from the right by $P_{23}$ results
in exchanging its second and third columns. Such an operation may
perfectly ruin the structure of this matrix. Only for rotations
$O_{23}\left(\pm{\pi/4}\right)$ by $\pm{\pi/4}$ in the $2$-$3$
plane or unitary matrices $U_{23}\left(\pm{\pi/4}\right)$
analogous to $U_{12}\left({\pi/ 4}\right)$ 
in eq.(\ref{eq:U12pi4})
the exchange of the second and third column can be
easily compensated for by some innocent change of conventions. But these are
exactly the matrices which describe the oscillations of atmospheric
neutrinos! If
\begin{equation}\label{VLUL}
V_L U_L^{-1} = O_{23}\left(\pm{\pi/4}\right)\quad{\rm\ or\ }\quad
V_L U_L^{-1} = U_{23}\left(\pm{\pi/4}\right),
\end{equation}
the resulting $U_{\rm MNS}$ can be cast in the form 
\begin{equation}
U_{\rm MNS} =
\left( \matrix{ 
1  & 0  &  0   \cr
0  & \cos\theta_{@}   & \sin\theta_{@}   \cr
0  & -\sin\theta_{@}   &  \cos\theta_{@}  \cr } \right)
\left( \matrix{ 
\cos \theta_\odot & \sin  \theta_\odot   &    0 \cr
-\sin\theta_\odot  & \cos\theta_\odot   &   0  \cr
0  &  0  &   1  \cr } \right)
\label{UMNS}
\end{equation}
by appropriate redefinitions of fields. This form
automatically guarantees that $U_{e3}$ is small. It is quite encouraging
that the condition (\ref{VLUL}) has been realized in many published
models, for a review see \cite{AFreview}. A particularly attractive
option is to assume 
\begin{equation}\label{UL1}
U_L \approx {\mathbf 1}
\end{equation}
and to attribute the whole unitary transformation in the $2$-$3$ plane to the
lepton sector. Then (\ref{VLUL}) can be obtained from the lopsided form of
$L$ \cite{ABB}. Eq.(\ref{UL1}) nicely agrees with the idea
that the analogous matrices for up and down quarks are
close to the unit matrix leading to the CKM matrix also close to one.

In grand unified theories like $SU_5$, the
mass matrices $L$ for leptons and $D$ for down quarks are closely
related:
\begin{equation}
L=D\TT
\end{equation}
This relation originates from the fact that in $SU_5$ the charged leptons 
are in fact closely related to 
charge conjugated fundamental fermions. So,
for $SO_{10}$ GUTs, eq.(\ref{UL1}) would imply that $V_R \approx
1$. However, in such theories there should be also a close relation
between $U_R$ and $V_L$. Comparing $V_L$ following from (\ref{VLUL}) for
$U_L=1$ and $U_R$ from (\ref{URsolved}) we do not find such a
relation. From (\ref{VLUL}) we can get
$U_{13}\left(\pm\pi/4\right)$
rather than $U_{23}\left(\pm\pi/4\right)$. So, we do not know
how to realize the scenario described in this paper in $SO_{10}$ unified 
theories.

After having observed the rather miraculous way the permutation $P_{23}$
leaves intact the pattern of mixing, it is easier to see why an
alternative matrix $\calN'$ for $(R)_{23}= {\cal O}(1)$
is not acceptable phenomenologically. We can take $R=P_{23}$, see
eq.(\ref{eq:P23}), as a representative for the whole
corresponding class of matrices. For this choice of $R$ we would obtain
\begin{equation}
\calN'=\frac{1}{M} U_L^{\rm T} \left(
\matrix{m_1^2 & 0       & 0 \cr
        0     & 0       & m_2 m_3 \cr
        0     & m_2 m_3 & 0}\right) U_L.
\end{equation}
The corresponding matrix $\calO'$, c.f. eq.(\ref{eq:18}) would be 
$\calO'=P_{13}U_{12}\left({\pi/4}\right)$, 
leading to the lepton mixing matrix
\begin{equation}\label{UMNSVLO'}
U_{\rm MNS}'=V_L U_L^{-1} P_{13} U_{12}\left({\pi}/{4}\right).
\end{equation}
As opposed to the expression (\ref{UMNSVUPUpi4}), the emerging mixing
pattern is hard to reconcile with data. Unlike that of the innocuous matrix
$P_{23}$ in (\ref{UMNSVUPUpi4}), the effect of the matrix $P_{13}$ in
(\ref{UMNSVLO'}) is deleterious to the structure of $V_L U_L^{-1}$ 
since it is now columns $1$ and $3$ that are swapped. To make
$U_{\rm MNS}'$ realistic, one would have to  resort to replacing the charged
lepton matrix $V_L$ with a much less plausible structure. 

\section{Towards a realistic model}

The central idea of the preceding sections is that the fundamental hierarchy
of the Dirac masses is drastically deformed by the matrix $R$. The form of
$R$ given in eq.(\ref{eq:R}) is particularly effective in reducing the
original hierarchy. For the same reason the role of the sub-leading terms
becomes much more prominent. In this section we show that realistic models
of lepton mixing and neutrino mass spectrum can be easily obtained when
these sub-leading terms are taken into account. Of course there is a
problem of the large number of free parameters in such purely
phenomenological models. Once again we face the fact that 
the amount of experimental information is very limited. 
So, one might argue that only more constrained models based on some 
underlying symmetry principle can be sufficiently constrained and 
predictive to go beyond a mere parametrisation of the data. 
We show that such a conclusion is too strong. In fact the
large hierarchy of the Dirac masses helps to reduce the number of relevant
sub-leading terms. Then the masses of the active neutrinos can be related
to $\dmatm$, $\dmsun$ and $\tan^2\theta_\odot$. These relations, in principle
at least, can provide an experimental test of the picture presented here.

Sub-leading terms are introduced as small off-diagonal elements in the
matrix $m^{(\nu)}$, see eq.(\ref{eq:mnu})\footnote{
An equivalent but less transparent way would be to preserve the diagonal
form of $m^{(\nu)}$  and add some sub-leading terms in $U_R$,
c.f. eq.(\ref{eq:N}). Small sub-leading terms in $U_L$ do not affect
the mass spectrum of the active neutrinos. Those terms may be very
important for the phenomenology of neutrinos because they can affect the
value of $U_{e3}$. However we have little to say about them.}. As we
want to preserve the hierarchy built in the Dirac mass spectrum we
assume that all non-zero off-diagonal elements of $m^{(\nu)}$ are small
and of order $m_1$. Then the structure of the mass operator $\calN$,
see eq.(\ref{eq:defN}), selects the matrix elements $m_{13}$ and $m_{12}$
whose contributions to $\calN$ are multiplied by $m_3$ and thus enhanced.
Let us assume that
\begin{equation}
m_{13} = a m_1  \quad {\rm and} \quad m_{12} = b m_1  .
\label{eq:m13}
\end{equation}
As explained above the contributions to $\calN$ from all other off-diagonal
elements of $m^{(\nu)}$ are even smaller and we neglect them in the
following discussion. The mass spectrum and the lepton mixing matrix
can be computed numerically for non-zero values of $a$ and $b$. It is
instructive, however, to consider the case $b=0$ which can be solved
analytically in terms of reasonably simple expressions. Moreover, it is
worth mentioning that the calculation of the leading contribution to 
the lepton mixing matrix, which was only sketched in Sec.3,
can be obtained from the following considerations for $a=0$.  
For $b=0$ 
\begin{equation}
m^{(\nu)}=\left(
\matrix{ m_1 & 0   &  a m_1 \cr
         0   & m_2 &  0     \cr
         0   & 0   & m_3}\right)
\end{equation}
and
\begin{equation}\label{P23MP23}
{\cal M} = P_{23} U_L^* {\calN} U_L^{-1} P_{23}
=\mu\left(\matrix{0&r&0\cr r& 2ar&0\cr 0&0&1}\right)
\end{equation}
with
$r = m_1 m_3/ m_2^2$ and $\mu = m_2^2 /M$,
where $P_{23}$ defined in eq.(\ref{eq:P23}) exchanges the second and 
third axes in the flavor space.
For $a>0$ the matrix ${\cal M}$ in (\ref{P23MP23}) is diagonalized by
\begin{equation}
U_{12}\TT {\cal M} U_{12} =\diag\left(\mu_1,\mu_2,\mu_3\right)
\label{eq:diagM}
\end{equation}
with
\begin{equation}\label{U12}
U_{12}\left( \alpha \right) =  
O_{12}\left( \alpha \right)\;\diag\left( i, 1, 1\right) =
\left( 
\matrix{ i\cos\alpha   &   \sin\alpha & 0 \cr
        -i\sin\alpha   &   \cos\alpha & 0 \cr
         0                    &  0                     & 1}
\right)
\end{equation}
and $\tan 2\alpha=1/a$,
which implies that 
\begin{equation}
\calO'=P_{23}U_{12}\left(\alpha\right).
\end{equation}
and the lepton mixing matrix is, cf.(\ref{UMNSVLO}),
\begin{equation}\label{UMNSVUPU}
U_{\rm MNS}=V_L U_L^{-1} P_{23} U_{12}\left( \alpha\right).
\end{equation}
It should be clear from the discussion in Sec.3 that the angle $\alpha$
in eq.(\ref{U12}) is equal to the solar mixing angle $\theta_\odot$.
The eigenvalues of the matrix ${\cal M}$ are
\begin{eqnarray}
\mu_1 &=& \mu r (\cosh t - \sinh t)  \cr
\mu_2 &=& \mu r (\cosh t + \sinh t)  \cr
\mu_3 &=& \mu 
\label{eq:mus}
\end{eqnarray}
with $a = \sinh t$. The ordering of the eigenvalues is such that e.g.
the second column of $U_{12}$ is equal to the normalized eigenvector 
of $\cal M$  for the eigenvalue $\mu_2$. This implies that
\begin{equation}
\cos\alpha = \left(\cosh t + \sinh t\right) \sin\alpha
\end{equation}
and
\begin{equation}
{\mu_1 / \mu_2} =
\tan^2\theta_\odot  .
\label{eq:38}
\end{equation}
The mass splittings are
\begin{equation}
\dmsun=\mu_1^2-\mu_2^2= 2\mu^2 r^2 \sinh2t,
\label{eq:39}
\end{equation}
\begin{equation}
 \dmatm\approx \mu^2,
\label{eq:40}
\end{equation}
which yields the formula for the parameter $\rho$, see eq.(\ref{eq:rho}),
\begin{equation}
\rho\approx 2r^2 \sinh2t.
\label{eq:rhosinh2t}
\end{equation}
Comparison with the allowed range for LMA MSW solar neutrino oscillations
\cite{SNO,phenosol} leads to the following ranges for the
values of $a$ and $r$:
\begin{equation}\label{eq:rangeofar}
 0.35 \le a \le 0.75\quad {\rm\ and\ }\quad  0.05 \le r \le 0.25 
\end{equation}
with the best fits corresponding to $a$ between 0.46 and 0.57
and $r$ between 0.09 and 0.10.\footnote{One might be tempted to assume in $N$
the same hierarchy as in $L$. Then the parameter $r=m_e
m_{\tau}/m_{\mu}^2$ is about $0.08$, which is in the range indicated
in eq.(\ref{eq:rangeofar}).
We are very much indebted to the Referee of PLB for pointing this
observation out to us.}

Eqs.(\ref{eq:38})-(\ref{eq:40}) lead to the following approximate 
formulae for the masses of the active neutrinos:
\begin{eqnarray}
\mu_1 &\approx& \sqrt{\dmsun}\; \tan^2\theta_\odot\;/ \;
\sqrt{1 - \tan^4\theta_\odot} \label{eq:41}\\
\mu_2 &\approx& \sqrt{\dmsun} \;/ \; \sqrt{1 - \tan^4\theta_\odot} 
\label{eq:42}\\
\mu_3 &\approx& \sqrt{\dmatm} \label{eq:43}
\end{eqnarray}
The errors due to approximations leading to these expressions are small 
for the interesting range of $a$ and $r$. 

Examine now the consequences of other  perturbations of the Dirac mass
matrix. Variation of the second important off-diagonal element,
$m_{12}$, shows that for non-zero values of the parameter $b$, see
eq.(\ref{eq:m13}), the mass relations (\ref{eq:41})-(\ref{eq:43})
are only weakly affected. In Fig.1 we show the test of the mass
relation (\ref{eq:38}) for a few values of $b$. The ratio plotted in Fig.1
is one if the formula (\ref{eq:38}) is exact. It is seen that corections
for non-zero values of $b$ are small. 
In Fig.2 we show the result of the analogous test of the formula 
(\ref{eq:41}). It is clear that it also works quite well.
\begin{figure}[ht]
\begin{center}
\epsfig{figure=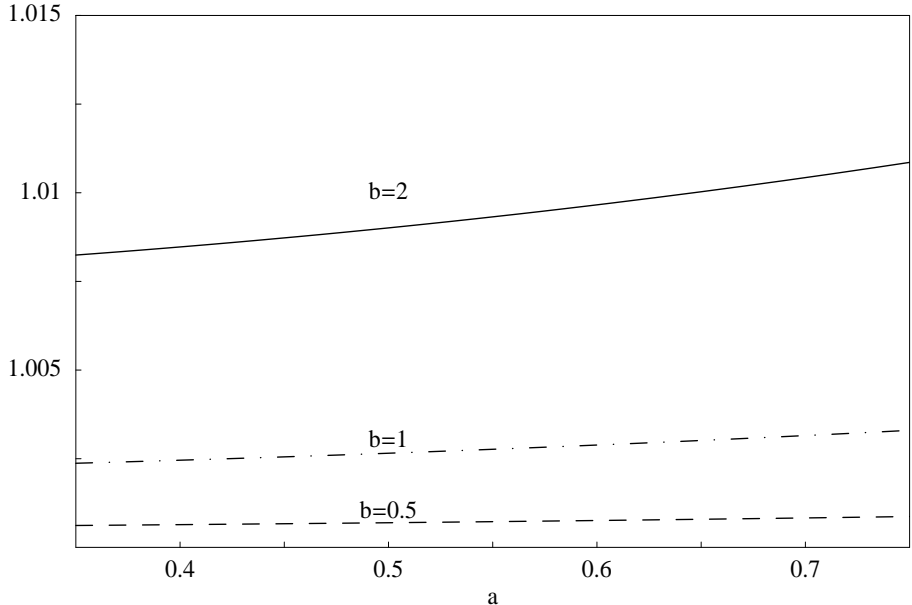,height=8cm,width=12cm}
\caption{
\label{Fig1}
Test of the relation {(\ref{eq:38})}
between the lighter neutrino masses and the tangent
of the sun mixing angle for non-zero values of $b$. The plot shows the
dependence of the ratio $\mu_1/(\mu_2 \tan^2\theta_\odot)$  on the
parameter $a$ in eq.(\ref{eq:m13})for different values of the
parameter $b$. For $b=0$ the value is exactly $1$.
}
\end{center}
\end{figure}
\begin{figure}[ht]
\begin{center}
\epsfig{figure=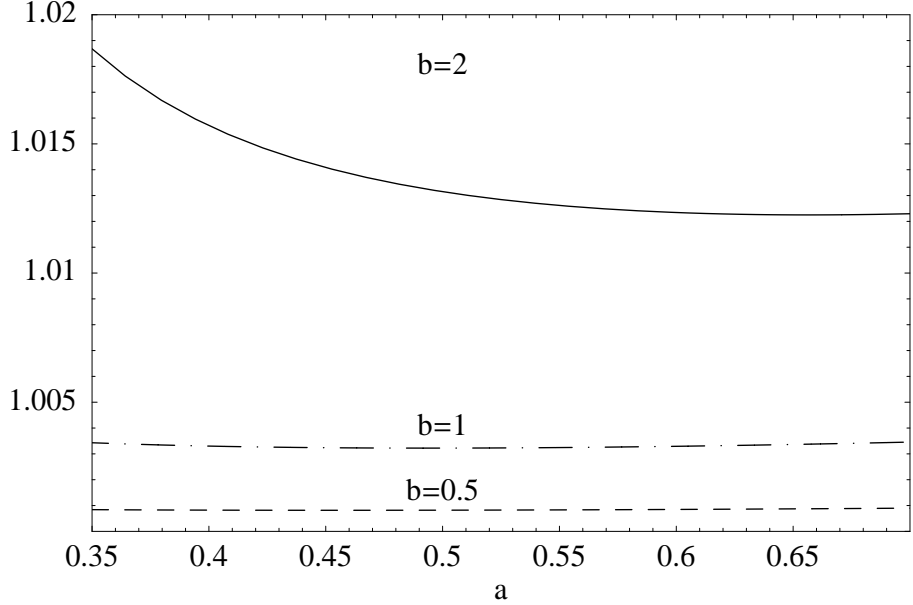,height=8cm,width=12cm}
\caption{
\label{Fig2}
Test of the formula {(\ref{eq:41})}
for $\mu_1$. Shown is the dependence of the ratio
$\mu_1 \sqrt{1-\tan^4\theta_\odot}\left/\left(\sqrt{\Delta m_\odot^2}
  \tan^2\theta_\odot\right)\right.$ 
on the parameter $a$ in eq.(\ref{eq:m13}) for different values of the
parameter $b$. 
}
\end{center}
\end{figure}
Non-zero $b$ affects the atmospheric mixing angle, yielding
roughly $\tan^2\theta_{@} \approx 1 + b$. Thus we must require that 
$|b| < 1$ if $0.95 \le \sin^2 2\theta_@ \le 1$ is to be obtained
without artificial conspiracy of contributions from the charged lepton
and neutrino sectors. We think that maximal or nearly maximal mixing
for the atmospheric neutrinos is an experimental fact which is too 
striking to assume that it is simply accidental and results from adding 
two contributions of similar size.  

It is worth noting that even though we have succeeded in keeping $U_{e3}$
zero, which is a non-trivial consequence of the structure of the lepton
and neutrino mass matrices, it is nevertheless possible to introduce
such a perturbation as to obtain any value of $U_{e3}$ being in
accordance with the present experimental data, i.e. $U_{e3} \le 0.1$.
This is effected by varying another off-diagonal element of the Dirac
matrix, $m_{21}$. One gets an approximate relation 
$U_{e3}\approx m_{21}/m_2$.

Let us close this section with two remarks on the numerical values
of the mass parameters in our picture. The mass of the lightest neutrino
mass eigenstate, see eq.(\ref{eq:41}), is about 3 meV for 
$\tan^2\theta_\odot\approx 0.4$. This mass range can be probed by the
10t version of the GENIUS project\cite{GENIUS} if the Majorana phases
are not too small and there are no strong cancellations between 
contributions to the mass parameter 
$\left< m_{\nu_e} \right>$. As a final remark 
let us note that the mass scale of the Majorana
masses is between $10^{10}$ and $10^{11}$ GeV if $m_2 \sim m_c$ 
is assumed. It has been pointed out in \cite{jez01} that 
this is exactly the range of Majorana masses which 
may be important for baryogenesis; 
see \cite{buchmuller} and references therein.

\section{Summary}
It has been shown that the Dirac masses of all fundamental fermions can
exhibit a strong hierarchy. A rather small hierarchy of the low energy 
neutrino masses is due to the symmetric unitary operator $R$ acting in 
the flavor space. 
This phenomenon results from the see-saw mechanism and the algebraic
structure of the dimension five effective mass operator ${\cal N}$ describing
the masses of the active neutrinos. In the leptonic sector there are two 
operators acting in the flavor space and observable at low energies:
the $\UMNS$ lepton mixing matrix and the matrix $R$. The matrix $R$ affects 
the form of $\UMNS$ and, moreover, also the low energy mass spectrum. The
form of $R$ proposed in \cite{j02} leads to a realistic description of 
neutrino mixing and masses. The mass of the heaviest neutrino is related 
to the mass scale $\sqrt{\dmatm}$ governing the oscillations of atmospheric 
neutrinos.
The masses of the two lighter neutrinos are related to $\sqrt{\dmsun}$,
the mass scale for oscillations of solar neutrinos and to the solar neutrino
mixing angle through the relation $\mu_1/\mu_2\approx \tan^2\theta_\odot$.
The structure of the lepton mixing matrix $\UMNS$ strongly suggests that
the maximal or nearly maximal mixing for the atmospheric neutrinos results
from the charged lepton sector. Small $U_{e3}$ is obtained as a consequence
of the model. The mass $\mu_1$ of the lightest neutrino mass eigenstate
is about 3 meV which can be observed by 10t GENIUS detector if Majorana 
phases are not too small and there are no strong cancellations between 
contributions to the mass parameter $\left< m_{\nu_e} \right>$.

\section{Acknowledgments}
MJ is very much indebted to Frans Klinkhamer for stimulating
and encouraging discussions. Many thanks are due to Guido Altarelli,
K.S. Babu, Andrzej Bia{\l}as, Ikaros Bigi, Wilfried Buchm\"uller, Ferruccio
Feruglio, Harald Fritzsch, Jean-Marc Gerard, Jack Gunion, Wolfgang
Hollik, Stanis{\l}aw Jadach, Hans Kuehn, Zoltan Kunszt, Paul
Langacker, Wiliam Marciano, Thomas Mannel, Rabindra Mohapatra, Holger Bech
Nielsen, Hans-Peter Nilles, Peter Minkowski, Stefan Pokorski, 
Leszek Roszkowski, Berthold Stech, Tsutomu Yanagida, Osamu Yasuda and
Kacper Zalewski for helpful comments, questions,   
remarks and suggestions. 

This work was done during our stay in the Institut f. Theoretische
Teilchenphysik, Universitaet Karlsruhe (TH).  
We would like to thank the Alexander-von-Humboldt Foundation for 
grants which made this possible. A warm atmosphere in TTP is gratefully
acknowledged. 

This work is also supported in part by the KBN grants 
5P03B09320 and 2P03B13622, and by 
the European Commission 5th Framework contract HPRN-CT-2000-00149.

\end{document}